\documentclass[runningheads]{cl2emult}

\usepackage{makeidx}  
\usepackage{graphicx} 
\usepackage{subeqnar} 
\usepackage{multicol} 
\usepackage{cropmark} 
\usepackage{eso}      
\makeindex            

\begin{document}
\title*{Analysis of CMB foregrounds using a database for Planck}
\toctitle{Analysis of CMB foregrounds using a database for Planck}
%
%
\titlerunning{Analysis of CMB foregrounds}
%
\author{G. Giardino\inst{1}
\and A. J. Banday\inst{2}
\and K. Bennett\inst{1}
\and P. Fosalba\inst{1}
\and K. M. G\'orski\inst{3}
\and W. O'Mullane\inst{1}
\and J. Tauber\inst{1}
\and C. Vuerli\inst{4}}
\authorrunning{Giovanna Giardino et al.}
%
%
\institute{Astrophsics Division -- Space Science Department of ESA, \\
        Noordwijk 2200 AG, The Netherlands
\and Max-Planck Institut f{\"u}r Astrophysik, Garching bei M\"{u}nchen,
                                             D-85741, Germany 
\and ESO, Garching bei M\"{u}nchen, D-85748, Germany
\and Osservatorio Astronomico di Trieste, 34131 Trieste, Italy}

\maketitle              

\begin{abstract}

\end{abstract}
Within the scope of the Planck IDIS (Integrated Data Information
System) project we have started to develop the data model for
time-ordered data and full-sky maps. The data model is part of the Data
Management Component (DMC), a software system designed according to a
three-tier architecture which allows complete separation between data
storage and processing.  The DMC is already being used for simulation
activities and the modeling of some foreground components.  We have
ingested several Galactic surveys into the database and used the
science data-access interface to process the data.  The data structure
for full-sky maps utilises the HEALPix tessellation of the sphere. We
have been able to obtain consistent measures of the angular power
spectrum of the Galactic radio continuum emission between 408~MHz and
2417~MHz.

\section{Introduction}
The ESA satellite Planck will survey the microwave sky with
unprecedented sensitivity from 30 to 850 GHz. These observations will
generate a large data set, with inhomogeneous data types, which will
need to be accessible to users spread across many institutes in Europe
and the US. How well it will be possible to mine this wealth of
information depends critically on the proper design of the Data
Management Component (DMC); that is on the efficiency and
user-friendliness of the suite of tools that will be used to store,
retrieve, process and query the data.

The IDIS \cite{idis} project is a collaboration among the two Planck
Data Processing Centres -- in particular OAT of Trieste, the Max-Planck
Institut f{\"u}r Astrophysik and the Astrophysics Division of ESTEC.
Within IDIS, we have started to develop the Planck DMC according to a
3-tier architecture.

\section{A 3-tier design for the Planck DMC}

\begin{figure}
\centering
\includegraphics[width=\textwidth,clip=]{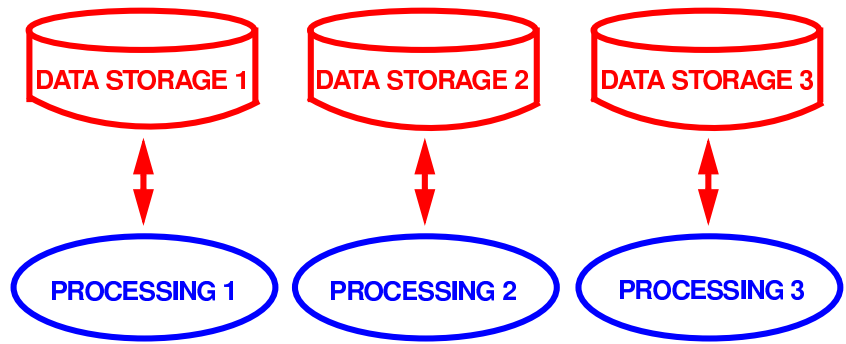}
\caption[]{In a 2-tier architecture changes to the storage system impact
 the processing system}
\label{fig:2tier}
\end{figure}

The two main logical partitions of data management are the data storage
system and its tools and the processing system and its tools. The
problem with this type of partitioning is that processing and storage
are not shielded from one another: changing storage system implies
making changes to some parts of the processing system (as illustrated
in Fig.~\ref{fig:2tier}). Moreover anyone concerned with the
development of the processing techniques has to be aware of storage
details, such as whether data are stored in files or in a database
system.  For a large project, involving hundreds of people
(scientists and software engineers) and lasting for over a decade, this
kind of approach is inefficient and can lead to a waste of resources.

\begin{figure}
\centering
\includegraphics[width=\textwidth]{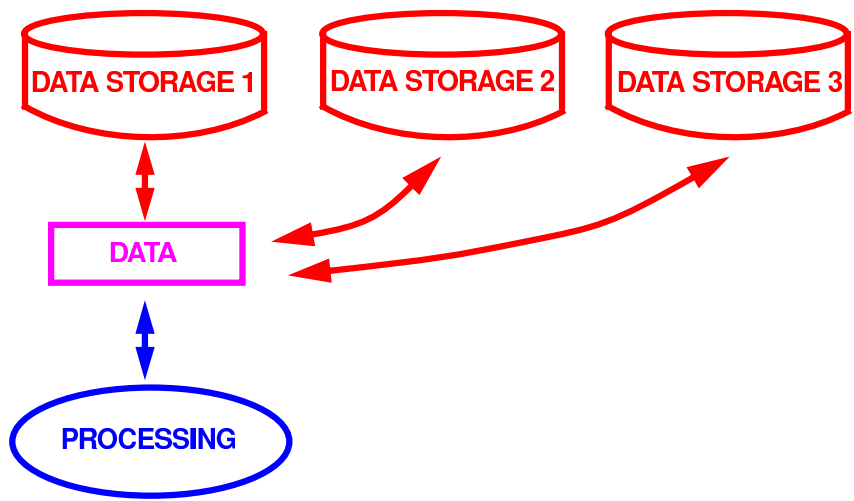}
\caption[]{In a 3-tier architecture Storage System and Processing System are
  de-coupled from each other: changes to the storage do not affect the
  processing pipelines. One can also envisage different storage systems
  being used in parallel.}
\label{fig:3tier}
\end{figure}

In a 3-tier design an additional layer is inserted between the data storage
and the processing layer. This is the data layer and in our case this
is an {\em abstract} layer.  All the processing is done using 
abstract objects or interfaces. Manipulating abstract objects rather
than real objects allows the user to develop programs which are more
closely related to the way one thinks about a problem rather than the
way a computer operates \cite{Gamma95}.  The extra tier keeps data and processing
completely separate from each other: changing the data storage system
does not have any implications for the processing
(Fig.~\ref{fig:3tier}).  This means, for instance, that if technology
evolves and a more efficient storage system becomes available, the
processing pipeline can be switched to access the data from the new
storage system, without any disruption.  In addition, thanks to this
extra layer, scientists can develop the processing algorithms without
having to worry about whether the data will be stored in files or in
databases.

The concept is very simple, but implementing this design requires extra
thought and some extra work in the initial phases of the project.
However, good design makes all the difference between allowing users
(the scientists) to efficiently mine the data or having them spending
hours debugging code.

We have started to develop the data access interfaces and the data
structures for Planck time-ordered data and Planck full-sky maps. The
latter uses the HEALPix tessellation of the sphere.  Implementation of
the system is done using an object oriented approach and the Java
programming language, which offers a natural way of implementing
abstract classes and interfaces.  To implement the database we decided
to use an object oriented database ({\em Objectivity}).  The DMC is
already being used for Planck simulation activities and the analysis of
some CMB foregrounds.

\section{The angular power spectrum of Galactic radio emission}

We have used this data management component and the database to measure
the angular power spectrum of Galactic radio emission using some of the
existing radio surveys. Galactic radio emission is a foreground signal
when observing the CMB.  The knowledge of the power spectra of the
foreground components is important in order to quantify the level of
contamination of the CMB observations at the different angular scales.
Improved modeling of the foregrounds also allows more realistic
simulations of the mission to be performed.

The Haslam survey at 408~MHz \cite{haslam}, the Reich \& Reich survey
at 1420~MHz \cite{reich} and the Jonas survey at 2326~MHz
\cite{Jonas98} were fed into the database.  The maps were
ingested raw, as they are publicly available. No de-striping process
was applied to the maps. The point sources were removed by median
filtering.  After point source removal, we could swiftly obtain a
measure of the angular power spectrum of the diffuse emission at the
different frequencies of the three maps.  

\begin{figure}
\centering
\includegraphics[width=8.0cm]{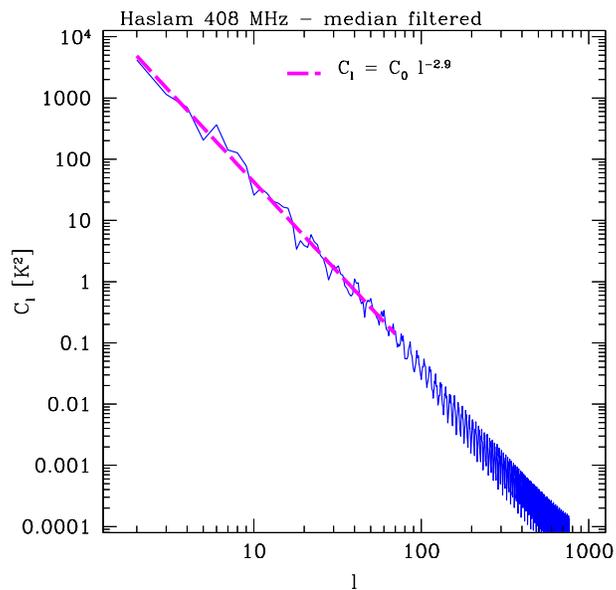}
\caption[]{The angular power spectrum of the Haslam survey at 408 MHz, for regions of the
  sky with $|b| > 20^{\circ}$. Points sources were removed by median
  filtering. The aliasing noise introduced by the Galactic plane cutoff
  at $ 20^{\circ}$ dominates over the steeply falling signal at $l >
  100$.}
\label{haslamAPS}
\end{figure}

\begin{figure}
\centering
\includegraphics[width=8.0cm]{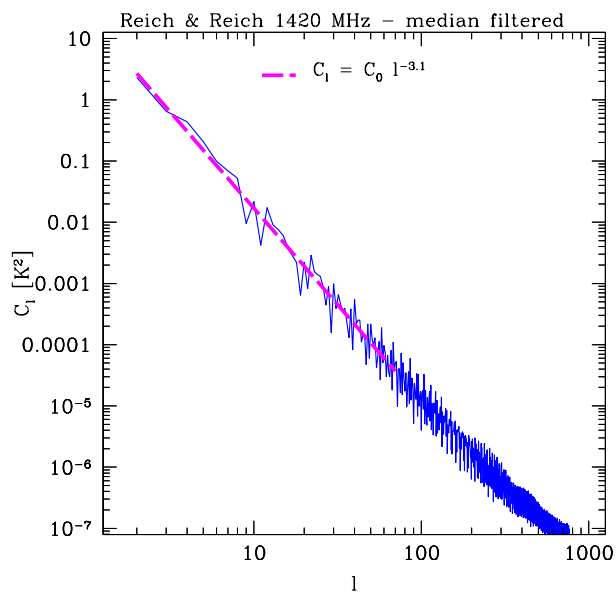}
\caption[]{The angular power spectrum of the Reich \& Reich survey at 1420 MHz,  for regions of the
sky with $|b| > 20^{\circ}$ (as per Fig.~3)}
\label{reichAPS}
\end{figure}

\begin{figure}
\centering
\includegraphics[width=8.0cm]{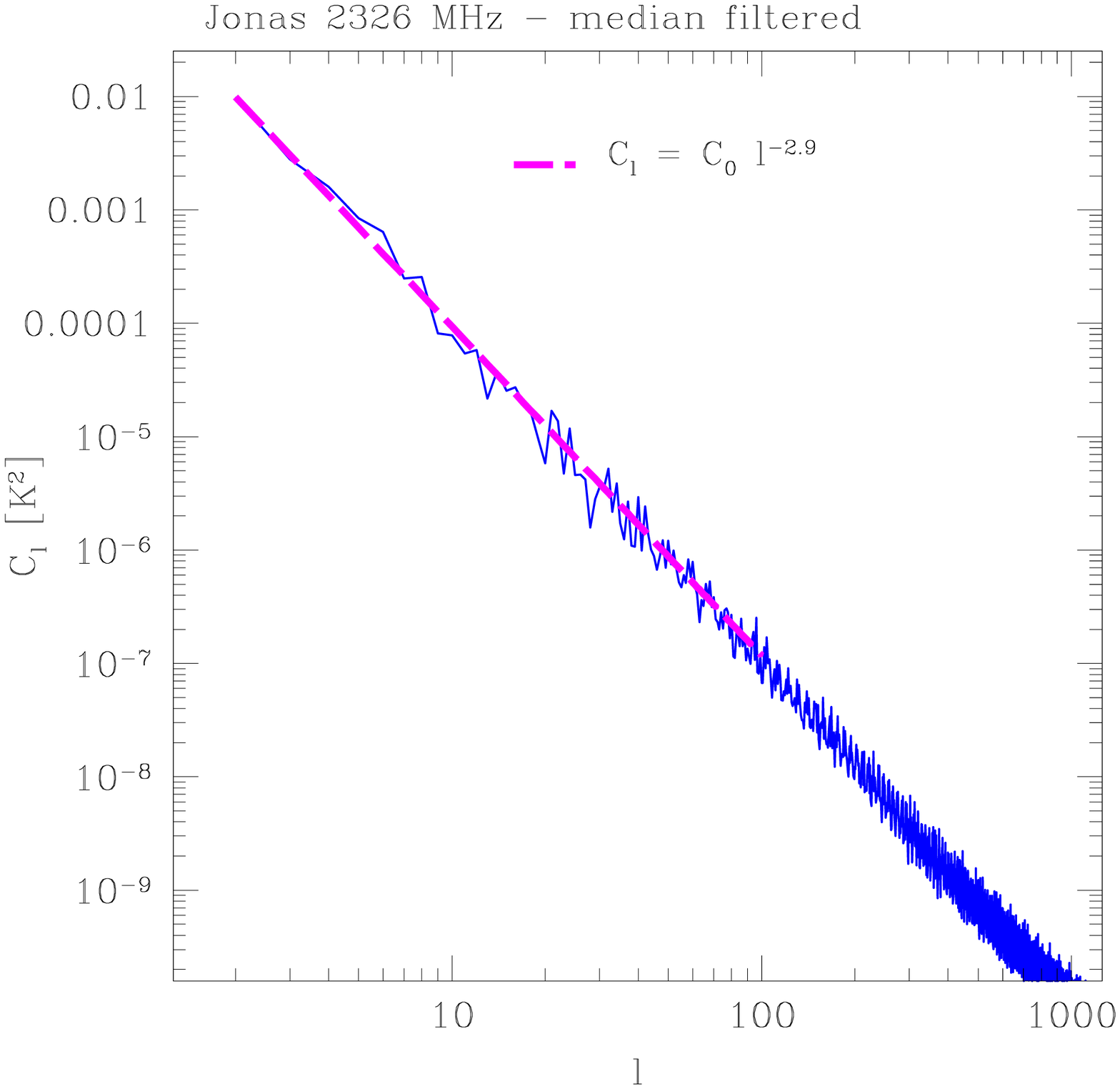}
\caption[]{The angular power spectrum of the Jonas survey at 2326 MHz, for regions of the
sky with $|b| > 20^{\circ}$ (as per Fig.~3).}
\label{jonasAPS}
\end{figure}

We model these spectra with a power law of the form $C_l \propto
l^{-\alpha}$ (see Giardino et al., in preparation -- for more details
about the analysis performed).  At high Galactic latitudes ($|{\rm b}|
> 20^{\circ}$) we derived a spectral index which is consistent for all
the 3 surveys and has an average value of $\alpha = 3.0 \pm 0.2$.  The
angular power spectra derived at high galactic latitude are reported in
Fig.~\ref{haslamAPS} for the Haslam map, in Fig.~\ref{reichAPS} for the
Reich \& Reich map and in Fig.~\ref{jonasAPS} for the Jonas map. The
best-fit power law spectra for each case are also shown. The derived
spectral indexes are summarized in Table~\ref{indexes}.

\begin{table}
\centering
\caption{Spectral index of the angular power spectrum of diffuse radio emission at high
  Galactic latitude for three available surveys. $\sigma_{\alpha}$ is
  the standard deviation of $\alpha$. The impact of incomplete sky
  coverage and the Galactic plane cutoff has been evaluated through
  Monte Carlo simulation.}
\renewcommand{\arraystretch}{1.4} \setlength\tabcolsep{5pt}
\begin{tabular}{llllll}
\hline\noalign{\smallskip}
Survey & $\nu$[MHz] & $\alpha$ & $\sigma_{\alpha}$ & $l$ range\\
\noalign{\smallskip}
\hline
\noalign{\smallskip}
Haslam & 408 & 2.94 & 0.09 & 2$-$70\\
Reich \& Reich & 1420 & 3.15 & 0.14 & 2$-$70\\
Jonas & 2326 & 2.92 & 0.07 & 2$-$100\\
\hline
\end{tabular}
\label{indexes}
\end{table}

\begin{figure}
\centering
\includegraphics[width=8.0cm]{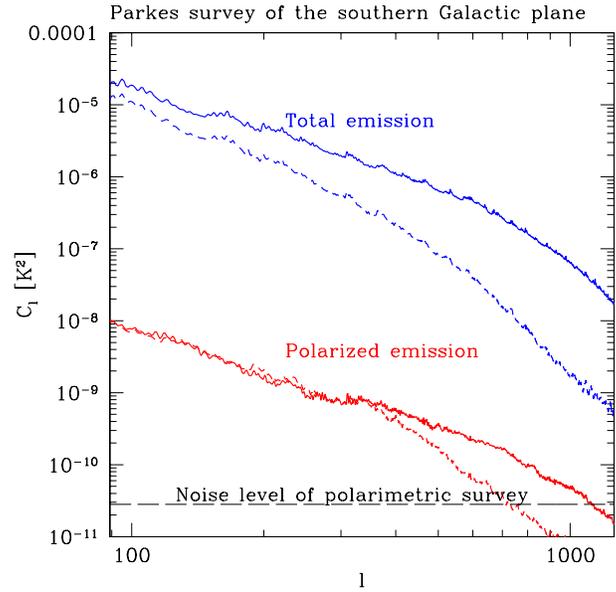}
\caption[]{The angular power spectrum of the Parkes survey of the
  southern Galactic plane at 2417 MHz. The angular power spectrum of
  the total emission  and the polarized fraction  are
  shown, before and after median filtering (solid and dashed line
  respectively)}
\label{duncan}
\end{figure}

The same analysis has also been applied to the Parkes polarimetric
survey of the Galactic Plane at 2417 MHz \cite{Duncan97} in order to
derive the angular power spectrum of the polarised emission.  From the
analysis of the raw data of the polarised component we have obtained a
spectral index of $\alpha = 1.9\pm 0.3$ in the multipole range $l \in
[100, 500]$ .  After median filtering the spectral index does not
change significantly in the $l$-range $100-300$\footnote{$l \leq 300$
  is the multipole range not affected by the median filter suppression
  of the high spatial-frequency signal}, while the spectral index of
the spectrum of the total intensity does (see Fig.~\ref{duncan}).  This
is expected since the contribution of point sources to the polarised
emission appears to be smaller than the point source contribution to
the total intensity.  Therefore, at 2.4 GHz and for regions at low
Galactic latitudes, the angular power spectrum of polarised diffuse
emission is significantly flatter than the angular power spectrum of
total diffuse emission.

\section{Conclusions}

This initial version of the Planck DMC allows maps from large sky
surveys to be handled efficiently and it can be used to perform
scientific data analysis. From the point of view of the client
(scientist) who is interested in data processing:
\begin{itemize}

\item the tool with its 3-tier design provides a convenient way to
  access the data. It allows the user to construct and
  handle ``scientific objects'' (such as maps or spectra) without having
  to understand the technicalities of data storage (formatting,
  optimization issues)
\item switching from a storage implementation which uses a file system
  on disk (e.g. a set of FITS files) or an object oriented database on
  a server is effortless 
\item the use of a data structure such as HEALPix has proved to be a
  crucial factor for the speed of operations involving Spherical
  Harmonic decomposition
\end{itemize}
From the point of view of the software engineer who develops the data
storage system:
\begin{itemize}

\item a database offers a way of handling and managing the data which
  is more powerful than a simple file system
\item from benchmarking tests, the use of an object oriented database in
  our case proved to be preferable to a relational database: for speed
  of data access and the option of storing objects of any given
  complexity (that is objects which refer to other objects, which in
  turn refer to other objects and so forth)
\end{itemize}

Using this version of the Planck DMC we have derived the spectral
indices of radio diffuse emission from the available large-sky surveys.
At high Galactic latitudes these are consistent with the spectral
indices derived previously from the Haslam map and the Reich \& Reich
map (\cite{Tegmark96}; \cite{bouchet96}; \cite{Bouchet99}), but are
more precise (Giardino et al., in preparation).

%

\clearpage
\addcontentsline{toc}{section}{Index}
\flushbottom
\printindex

\end{document}